\documentclass[twocolumn,showpacs]{revtex4}

\usepackage[dvips]{graphicx}
\usepackage{color}

\begin{document}

\title{Atomic Quantum Memory for Photonic Qubits
via Scattering in Cavity QED}
\date{\today}
\author{Hiroyuki Yamada}
\author{Katsuji Yamamoto}
\affiliation{Department of Nuclear Engineering,
Kyoto University, Kyoto 606-8501, Japan
}

\begin{abstract}
We investigate a scheme of atomic quantum memory
to store photonic qubits in cavity QED.
This is motivated on the recent observation that
the quantum-state swapping between a single-photon pulse and
a $ \Lambda $-type atom trapped in a cavity
is ideally realized via scattering for some specific case
in the strong coupling cavity regime
[T.~W.~Chen, C.~K.~Law, and P.~T.~Leung,
Phys. Rev. A {\bf 69}, 063810 (2004)].
We derive a simple formula for calculating the fidelity
of this atom-photon swapping for quantum memory.
We further propose a feasible method
which implements conditionally the quantum memory operation
with the fidelity of almost unity
even if the atom-photon coupling is not so strong.
This method can also be applied to store a photonic entanglement
in spatially separated atomic quantum memories.
\end{abstract}

\pacs{42.50.Pq, 42.50.Dv, 03.67.Hk}

\maketitle

Combined systems of atoms and photons have been studied extensively
to construct quite promising and efficient quantum networks
for information processing and communication
\cite{qnet}.
In these quantum networks, quantum-state transfer
between photons and atoms (matter)
and storage of quantum states are particularly important.
Then, numerous methods to implement the quantum-state transfer
and quantum memory have been proposed and investigated
in various manners
\cite{qtrqm-1,qtrqm-2,qtrqm-3,qtrqm-4,qtrqm-5,qtrqm-6,qtrqm-7,
qtrqm-8,qtrqm-9,qtrqm-10,qtrqm-11}.
The cavity QED is among the promising schemes
to realize such quantum-state operations,
which utilizes strong interaction between single atoms and photons
inside cavities
\cite{CQED}.
Specifically, quantum-state transfer and manipulation
are made between a single atom and a single-photon pulse
through a scattering in an optical cavity.
In this Letter, we investigate a scheme of atomic quantum memory
to store photonic qubits in cavity QED.
This is motivated on the recent observation that
the quantum-state swapping between a single-photon pulse and
a $ \Lambda $-type atom trapped in a cavity
is ideally realized via scattering for some specific case
in the strong coupling cavity regime producing the maximal phase shift
\cite{CLL-2004}.

We consider a one-dimensional cavity bounded by two mirrors,
one of which is perfectly reflecting
while the other is partially transparent.
The electromagnetic field is expanded in terms of the continuous modes
with wave number $ k $, which range over the inside of cavity
through the outside free space
\cite{CLL-2004}.
A photonic qubit is encoded in the polarization states
$ | k_L \rangle $ and $ | k_R \rangle $ of single-photon pulse as
\begin{eqnarray}
| \phi_{\rm p} \rangle
&=& c_L | {\bar k}_L \rangle + c_R | {\bar k}_R \rangle ,
\label{eqn:phi-p}
\\
| {\bar k}_{L,R} \rangle
&=& \int_{- \infty}^{\infty} dk f(k) e^{-ikt} | k_{L,R} \rangle ,
\end{eqnarray}
where $ f(k) $ is the normalized spectral amplitude,
and $ e^{-ikt} $ represents the asymptotic temporal evolution
($ c = \hbar = 1 $ unit).
On the other hand, a $ \Lambda $-type atom is trapped
inside the cavity, which has two degenerate ground states
$ | L \rangle $ and $ | R \rangle $ and an excited state $ | e \rangle $.
Then, an atomic qubit is encoded in the degenerate ground states as
\begin{eqnarray}
| \psi_{\rm a} \rangle = a_L | L \rangle + a_R | R \rangle .
\label{eqn:psi-a}
\end{eqnarray}

The polarization states $ | k_L \rangle $ and $ | k_R \rangle $
are coupled, respectively, to the transitions
$ | L \rangle - | e \rangle $ and $ | R \rangle - | e \rangle $
of frequency $ \omega_e $ in the cavity with the dipole couplings
\begin{eqnarray}
g_{L,R} (k) = \lambda_{L,R} {\sqrt{\kappa / \pi}}
e^{i \theta_{L,R}} / ( k - k_c + i \kappa ) ,
\end{eqnarray}
where $ k_c $ is the resonant frequency of the cavity,
$ \kappa $ is the leakage rate of the cavity,
$ \lambda_L $ and $ \lambda_R $ represent the normalized coupling strengths,
and $ \theta_L $ and $ \theta_R $ are the phase angles
from the dipole transition matrix elements.
The atom-photon scattering then takes place through these couplings,
and the transformation of the atom-photon states is induced
asymptotically as
\begin{eqnarray}
{\cal T} | L k_L \rangle
&=& T_{LL} (k) | L k_L \rangle + T_{RL} (k) | R k_R \rangle ,
\nonumber \\
{\cal T} | R k_R \rangle
&=& T_{LR} (k) | L k_L \rangle + T_{RR} (k) | R k_R \rangle ,
\nonumber \\
{\cal T} | L k_R \rangle
&=& | L k_R \rangle , \
{\cal T} | R k_L \rangle = | R k_L \rangle ,
\label{eqn:transform}
\end{eqnarray}
where the basis states are taken as
$ | L k_L \rangle \equiv | L \rangle | k_L \rangle $, and so on.
The scattering matrix elements are calculated explicitly
in Ref. \cite{CLL-2004} as
\begin{eqnarray}
&& T_{LL} (k) = e^{i \phi_s (k)}| \xi_L (k) |^2+| \xi_R (k) |^2 ,
\nonumber \\
&& T_{RR} (k) = e^{i \phi_s (k)}| \xi_R (k) |^2+| \xi_L (k) |^2 ,
\nonumber \\
&& T_{LR} (k) = \xi^*_L (k) \xi_R (k) ( e^{i \phi_s (k)} - 1 ) ,
\nonumber \\
&& T_{RL} (k) = e^{2i ( \theta_L - \theta_R )}
\xi^*_L (k) \xi_R (k) ( e^{i \phi_s (k)} - 1 ) ,
\end{eqnarray}
where $ \xi_{L,R} (k)
\equiv g_{L,R} (k) / \sqrt{| g_L (k) |^2 + | g_R(k) |^2} $.
Here, the bright state acquires a phase shift $ \phi_s(k) $ via scattering.
This linear transformation $ {\cal T} $ with a complex $ \phi_s(k) $
is generally non-unitary due to the loss with a rate $ \gamma $
induced by the spontaneous emission into the environment.

\begin{flushleft}
{\it Swapping for qubit memory}
\end{flushleft}

It is observed \cite{CLL-2004}
that the quantum-state swapping between the atom and photon
can be made ideally via scattering in the specific case
of the $ \Lambda $-type atom with equal but opposite
dipole matrix elements, i.e., $ \lambda_L = \lambda_R = \lambda $
and $ e^{i( \theta_L - \theta_R )} = - 1 $, which provides
\begin{eqnarray}
g_L (k) = - g_R (k) .
\label{eqn:gL-gR}
\end{eqnarray}
For example, we may take the D1 line of sodium with
$ | L \rangle = | F = 1 , m_F = -1 \rangle $,
$ | R \rangle = | F = 1 , m_F = 1 \rangle $,
$ | e \rangle = | F = 1 , m_F = 0 \rangle $.
In fact, with the maximal phase shift
$ e^{i \phi_s ( k_c )} = - 1 $ at the resonance
for $ \kappa \gamma / \lambda^2 \rightarrow 0 $,
we have the scattering matrix elements as
\begin{eqnarray}
T_{LR}( k_c ) = T_{RL}( k_c ) = 1 , \
T_{LL}( k_c ) = T_{RR}( k_c ) = 0 .
\end{eqnarray}
Then, the atom-photon swapping is obtained as
\begin{eqnarray}
| \Phi_{\rm in}^{(k)} \rangle
&=& ( a_L | L \rangle + a_R | R \rangle )
\otimes ( c_L | k_L \rangle + c_R | k_R \rangle ) ,
\label{eqn:Phi-in}
\\
& \Updownarrow &
\nonumber \\
| \Phi_{\rm swap}^{(k)} \rangle
&=& ( c_R | L \rangle + c_L | R \rangle )
\otimes ( a_R | k_L \rangle + a_L | k_R \rangle ) .
\label{eqn:Phi-swap}
\end{eqnarray}
We here note that this swapping is made in a reversible way
via scattering.
Hence it can be applied to implement an atomic quantum memory
for the storage of unknown photonic qubits of polarization.
The input photonic qubit is stored ({\it written}) via scattering,
and it is retrieved ({\it read}) by injecting another single-photon pulse.

We now evaluate the fidelity of this atom-photon swapping
for the specific case of $ g_{L} (k) = - g_{R} (k) $
providing $ T_{LR}(k) = T_{RL}(k) $ and $ T_{LL}(k) = T_{RR}(k)
= 1 - T_{LR}(k) $.
Arbitrary atomic and photonic qubits
in Eqs. (\ref{eqn:phi-p}), (\ref{eqn:psi-a}) and (\ref{eqn:Phi-in})
may be taken as the initial state $ | \Phi_{\rm in} \rangle $.
Then, the density operator of the output state via scattering
is given by
\begin{eqnarray}
\rho_{\rm out} &=& | \Phi_{\rm out} \rangle \langle \Phi_{\rm out} |
+ ( 1 - \langle \Phi_{\rm out} | \Phi_{\rm out} \rangle )
| 0 \rangle \langle 0 | ,
\\
| \Phi_{\rm out} \rangle
&=& {\cal T} | \Phi_{\rm in} \rangle
= \int_{- \infty}^{\infty} dk f(k) e^{-ikt}
| \Phi_{\rm out}^{(k)} \rangle ,
\\
| \Phi_{\rm out}^{(k)} \rangle
&=& T_{LR}(k) | \Phi_{\rm swap}^{(k)} \rangle
+ T_{LL}(k) | \Phi_{\rm in}^{(k)} \rangle .
\end{eqnarray}
Here, the term of $ | 0 \rangle \langle 0 | $
represents the loss due to the spontaneous emission
with $ {\rm Tr} \rho_{\rm out} = 1 $.
The output photon will eventually be absorbed by matter.
Then, by taking the trace over the photon states
the fidelity to obtain the desired atomic state
$ | \psi_{\rm swap} \rangle = c_R | L \rangle + c_L | R \rangle $
is given by
\begin{eqnarray}
F = \left[
\langle \psi_{\rm swap} | {\rm Tr}_{(k)}
\left[ | \Phi_{\rm out}^{(k)} \rangle \langle \Phi_{\rm out}^{(k)} | \right]
| \psi_{\rm swap} \rangle \right]_f ,
\end{eqnarray}
where $ {\rm Tr}_{(k)} [ \rho ] \equiv \langle k_L | \rho | k_L \rangle
+ \langle k_R | \rho | k_R \rangle $,
and the average of any function $ G(k) $ of $ k $
with the weight $ | f(k) |^2 $ is denoted as
\begin{eqnarray}
[ G ]_f \equiv \int_{- \infty}^{\infty} dk | f(k) |^2 G(k) .
\end{eqnarray}
This fidelity is calculated by considering
$ T_{LR} (k) + T_{LL} (k) = 1 $ as
\begin{eqnarray}
F(D) = F(0) + [ 1 -  F(0) ] D
\end{eqnarray}
with $ D
= | \langle \psi_{\rm swap} | \psi_{\rm a} \rangle |^2
= | c_R^* a_L + c_L^* a_R |^2 $ ($ 0 \leq D \leq 1 $).
Hence, we may take the fidelity of swapping as
\begin{eqnarray}
F_{\rm swap}
= F(0) = \left[ | T_{LR}(k) |^2 \right]_f
\end{eqnarray}
irrespective to the choice of initial state.
Here it is desired for optimizing the quantum-state transfer via swapping
that the spectral width $ \kappa_{\rm p} $ of the photon pulse with $ f(k) $
should be made sufficiently smaller than the cavity leakage rate $ \kappa $,
as discussed in Ref. \cite{CLL-2004}.
Then, we have the quite high fidelity
$ F_{\rm swap} \simeq | T_{LR} ( k_c ) |^2 \simeq 1 $
with $ \kappa_{\rm p} \ll \kappa $
in the strong coupling regime
$ \kappa \gamma / \lambda^2 \ll 1 $.

Numerically, by using the formula for the phase shift $ \phi_s (k) $
\cite{CLL-2004}
we obtain $ F_{\rm swap}({\rm Gaussian}) = 0.975 $
typically with $ \lambda = 5 \kappa $, $ \gamma = 0.5 \kappa $
($ \kappa \gamma / \lambda^2 = 0.02 $) and $ \omega_e = k_c $
for the Gaussian $ | f(k) |^2
= \exp [ - ( k - k_c )^2 / \kappa_{\rm p}^2 ]/( \pi^{1/2} \kappa_{\rm p} ) $
with $ \kappa_{\rm p} = 0.1 \kappa $.
We also obtain
$ F_{\rm swap}({\rm Lorentzian}) = 0.887 (0.960) $
with $ \lambda = 5 \kappa $ and $ \gamma = 0.5 \kappa $
for the Lorentzian $ | f(k) |^2
= ( \kappa_{\rm p} / \pi )[ ( k - k_c)^2 + \kappa_{\rm p}^2 ]^{-1} $
with $ \kappa_{\rm p} = 0.1 \kappa ( 0.02 \kappa )$.
The  atomic detuning does not provide a significant effect
on the fidelity for $ | \omega_e - k_c | \lesssim \gamma \sim \kappa $.

\begin{flushleft}
{\it Storage and retrieval with conditional measurements}
\end{flushleft}

We next consider the sequence of storage and retrieval of photonic qubit,
which may appear somewhat different from simply repeating
twice the atom-photon swapping.
To be general, we relax the condition (\ref{eqn:gL-gR})
for the $ \Lambda $-type atom,
allowing different $ g_L (k) $ and $ g_R (k) $.
In this situation, as seen below, even for a not so strong
atom-photon coupling the almost faithful quantum memory operation
can be achieved conditionally
by making some projective measurements.
Specifically, the initial state is taken as
\begin{eqnarray}
| \Phi_{\rm in} \rangle
= | R \rangle | \phi_{\rm p1} \rangle | {\bar k}_R^\prime \rangle
= | R \rangle
( c_L | {\bar k}_L \rangle + c_R | {\bar k}_R \rangle )
| {\bar k}_R^\prime \rangle . \
\end{eqnarray}
Here, $ | \phi_{\rm p1} \rangle $ is the photonic qubit
to be stored and then retrieved.
The atomic state is initially prepared to be $ | R \rangle $,
and the second photon pulse of $ | {\bar k}_R^\prime \rangle $
is injected after a time delay $ \tau $
($ \gg \kappa_{\rm p}^{-1} \gg \kappa^{-1} $) to retrieve the stored qubit.
We take for definiteness the same profile $ f(k) $
for the two photon pulses, though this choice is not essential.

After the scattering of the first photon pulse with the atom,
the detection of the polarization ``$ L $" is made
on the output photon.
This polarization detection is represented
by a positive operator valued measure
\begin{eqnarray}
\Pi ( k_L ) = \int_{- \infty}^{\infty} dk \eta (k)
| k_L \rangle \langle k_L |
\end{eqnarray}
with the quantum efficiency $ 0 < \eta (k) \leq 1 $.
[The dark count is neglected here
since it can actually be made rather small.
The terms of more than one photon states
may also be discarded effectively in $ \Pi ( k_L ) $
in the present process involving a single atom and photon.]
Then, the resultant state is given by
\begin{eqnarray}
\rho_1 &=& P( k_L )^{-1} {\rm Tr}_{\rm p1} [ \Pi ( k_L )
{\cal T}_1 | \Phi_{\rm in} \rangle
\langle \Phi_{\rm in} | {\cal T}_1^\dagger ]
\nonumber \\
&=& P( k_L )^{-1} \int_{- \infty}^{\infty} dk \eta (k) | f(k) |^2
| \Phi_1^{(k)} \rangle \langle \Phi_1^{(k)} | ,
\end{eqnarray}
where
\begin{eqnarray}
| \Phi_1^{(k)} \rangle
= [ T_{LR}(k) c_R | L \rangle + c_L | R \rangle ]
| {\bar k}_R^\prime \rangle
\label{eqn:Phi-1}
\end{eqnarray}
by applying Eq. (\ref{eqn:transform}) for $ {\cal T}_1 $.
It is noticed in Eq. (\ref{eqn:Phi-1}) that the initial photonic qubit
is transferred to the atomic qubit with slight modification
by the factor $ T_{LR}(k) $.
The loss term of $ | 0 \rangle \langle 0 | $
is projected out by the photon detection even with $ \eta (k) < 1 $
(and the negligible dark count).
The success probability of the photon detection
is given by $ P( k_L ) = [ \eta (k)
\langle \Phi_1^{(k)} | \Phi_1^{(k)} \rangle ]_f $,
providing the normalization $ {\rm Tr}_{\rm ap2} \rho_1 = 1 $.

The retrieval of the photonic qubit is implemented
by the scattering of the second photon pulse
followed by the conditional detection of the atomic state $ | L \rangle $.
The resultant output state is given by
\begin{eqnarray}
\rho_{\rm out}
&=& P(L)^{-1} \langle L | {\cal T}_2 \rho_1 {\cal T}_2^\dagger | L \rangle
\nonumber \\
&=&
\frac{| T_{LR}( k_c ) |^2}{P( k_L ) P(L)}
\nonumber \\
&{}& \times
\int_{- \infty}^{\infty} dk
\eta (k) | f(k) |^2 
| \phi_{\rm out}^{(k)} \rangle \langle \phi_{\rm out}^{(k)} | ,
\label{eqn:rho-out}
\end{eqnarray}
where
\begin{eqnarray}
| \phi_{\rm out}^{(k)} \rangle
&=& \int_{- \infty}^{\infty} dk^\prime
f( k^\prime ) e^{-ik^\prime ( t - \tau )}
\nonumber \\
&{}& \times [ r_{LR}(k) c_R | k^\prime_R \rangle
+ r_{LR}( k^\prime ) c_L | k^\prime_L \rangle ]
\label{eqn:phi-out}
\end{eqnarray}
with
\begin{eqnarray}
r_{LR}(k) & \equiv & T_{LR} (k) / T_{LR} ( k_c ) .
\label{eqn:rLR}
\end{eqnarray}
It is found in Eq. (\ref{eqn:phi-out}) that
with $ T_{LR} (k) \approx T_{LR} ( k_c ) $, i.e.,
$ r_{LR}(k) \approx 1 $ in the vicinity of resonance
$ | k - k_c | \lesssim \kappa_{\rm p} \ll \kappa $
the output state is very closed to the desired photon state as retrieval:
\begin{eqnarray}
| \phi_{\rm out}^{(k)} \rangle \approx | \phi_{\rm p2} \rangle
= 
( c_L | {\bar k}_L^\prime \rangle + c_R | {\bar k}_R^\prime \rangle ) .
\label{eqn:phi-p2}
\end{eqnarray}
The net success probability of the storage and retrieval,
providing the normalization $ {\rm Tr}_{\rm p2} \rho_{\rm out} = 1 $,
is given by
\begin{eqnarray}
P(L)P( k_L ) = | T_{LR}( k_c ) |^2
\left[ \eta (k)
\langle \phi_{\rm out}^{(k)} | \phi_{\rm out}^{(k)} \rangle \right]_f .
\label{eqn:P-net}
\end{eqnarray}
Then, by considering Eqs. (\ref{eqn:rho-out})--(\ref{eqn:phi-p2})
the fidelity for this operation of storage and retrieval
is evaluated as
\begin{eqnarray}
F ( {\rm p1} \rightarrow {\rm a} \rightarrow {\rm p2} )
= \frac{\left[ \eta (k)
| \langle \phi_{\rm p2} | \phi_{\rm out}^{(k)} \rangle |^2
\right]_f}
{\left[ \eta (k)
\langle \phi_{\rm out}^{(k)} | \phi_{\rm out}^{(k)} \rangle
\right]_f} .
\label{eqn:Fp1ap2}
\end{eqnarray}

As seen in Eq. (\ref{eqn:phi-p2}),
we can obtain the fidelity of almost unity with $ r_{LR}(k) \approx 1 $
for $ | k - k_c | \lesssim \kappa_{\rm p} \ll \kappa $,
even if the atom-photon coupling is not so strong
as $ \lambda_{L,R} \sim \kappa \sim \gamma $
to give $ | T_{LR}( k_c ) |^2 \sim 0.1 $.
Specifically, by considering that the quantum efficiency
may be constant as $ \eta (k) = \eta $ in the vicinity of resonance,
the fidelity is calculated
as $ F ( {\rm p1} \rightarrow {\rm a} \rightarrow {\rm p2} )
= F_{\rm qm} + [ 1 - F_{\rm qm} ] | c_L |^4 $
depending on the initial $ | \phi_{\rm p1} \rangle $.
Then, the fidelity of this quantum memory operation is given by
\begin{eqnarray}
F_{\rm qm} = | [ r_{LR} ]_f |^2 / [ | r_{LR} |^2 ]_f \simeq 1 \
( \kappa_{\rm p} \ll \kappa ) .
\label{eqn:F-qm}
\end{eqnarray}
This is a quite remarkable feature of the present conditional scheme
for quantum memory, without requiring the specific $ \Lambda $-type atom
satisfying the condition (\ref{eqn:gL-gR}) on the dipole couplings.
Although some modification is made on the stored atomic state,
as seen in Eq. (\ref{eqn:Phi-1}),
it is nearly compensated by the read-out process,
as seen in Eq. (\ref{eqn:phi-out}),
realizing the almost faithful retrieval of the initial photonic qubit.
The trade-off of the success probability is instead
paid to obtain the high fidelity for the general case.

Numerically, we have estimates of the fidelity
and the net success probability
as $ F_{\rm qm} = {\mbox{0.995, 0.994, 0.999}} $,
and $ P(L) P( k_L ) = ({\mbox{0.975, 0.634, 0.248}}) \eta $,
for $ \lambda_L = \lambda_R = ({\mbox{5 , 1 , 0.5}}) \kappa $, respectively
with $ \gamma = 0.5 \kappa $, $ \omega_e = k_c $
and $ \kappa_{\rm p} = 0.1 \kappa $ for the Gaussian form.
Similar estimates are also obtained for the Lorentzian form
with $ \kappa_{\rm p} = 0.01 \kappa $.
Therefore, the quite high fidelity $ F_{\rm qm} $ is really obtained
almost independently of the atom-photon coupling
$ \lambda_{L,R} / \kappa \gtrsim 0.5 $.

It should be mentioned that the atomic detection of $ | L \rangle $
can be implemented by injecting the third photon pulse.
Specifically, by injecting the photon
of $ | {\bar k_L}^{\prime \prime} \rangle $
followed by the polarization detection $ \Pi ( k_R^{\prime \prime} ) $
on the output photon,
the $ | L \rangle $ component of the initial atomic state
is transformed to $ | R \rangle $ via scattering
with the success probability $ \approx \eta | T_{RL}( k_c ) |^2 $
while the $ | R \rangle $ one is projected out.

In a feasible experiment, a sufficiently weak coherent light
of $ | \alpha \rangle $ may be used as an actual single-photon source,
though the success probability becomes rather small
proportional to $ | \alpha |^2 $.
The vacuum contribution is projected out conditionally
by the detection of the output photon.
The contributions of more than one photon states
are small enough for $ | \alpha |^2 \ll 1 $.

\begin{flushleft}
{\it Storage of 2-qubit entanglement}
\end{flushleft}

We can see that the quantum-state transfer via scattering
can also be applied to the storage of 2-qubit entanglement.
We prepare two atomic memories
and a polarization-entangled pair of photon pulses.
Each photon pulse is scattered with the atom inside the respective cavity.
In this situation, particularly for the ideal case
of $ T_{LR} = T_{RL} = 1 $ and $ T_{LL} = T_{RR} = 0 $,
we obtain the swapping between the generic states
of atom pair and photon pair,
which may be either entangled or separable, as
\begin{eqnarray}
\left( \begin{array}{c} a_{LL} \\ a_{RR} \\
a_{LR} \\ a_{RL} \end{array} \right)_{\rm a}
\otimes
\left( \begin{array}{c} c_{LL} \\ c_{RR} \\
c_{LR} \\ c_{RL} \end{array} \right)_{\rm p}
\Leftrightarrow
\left( \begin{array}{c} c_{RR} \\ c_{LL} \\
c_{RL} \\ c_{LR} \end{array} \right)_{\rm a}
\otimes
\left( \begin{array}{c} a_{RR} \\ a_{LL} \\
a_{RL} \\ a_{LR} \end{array} \right)_{\rm p} ,
\nonumber \\
{}
\end{eqnarray}
where the basis states are taken as
for the atom pair ``a"
$ ( | L L \rangle , | R R \rangle , | L R \rangle , | R L \rangle ) $
and for the photon pair ``p"
$ ( | k_L k_L \rangle , | k_R k_R \rangle ,
| k_L k_R \rangle , | k_R k_L \rangle ) $.

This sort of 2-qubit transfer can readily be applied
to the storage of photonic polarization entanglement as
\begin{eqnarray}
c_{LR} | {\bar k}_L \rangle | {\bar k}_R^\prime \rangle
+ c_{RL} | {\bar k}_R \rangle | {\bar k}_L^\prime \rangle
\Rightarrow
c_{RL} | L R \rangle + c_{LR} | R L \rangle .
\end{eqnarray}
Specifically, by taking the initial state
\begin{eqnarray}
| \Phi_{\rm in} \rangle
= | R R \rangle
( c_{LR} | {\bar k}_L \rangle | {\bar k}_R^\prime \rangle
+ c_{RL} | {\bar k}_R \rangle | {\bar k}_L^\prime \rangle ) ,
\end{eqnarray}
we obtain the $ ( k k^\prime ) $-component of the output state
via scatterings in the cavities 1 and 2 as
\begin{eqnarray}
| \Phi_{\rm out}^{( k k^\prime )} \rangle
&=& [ T_{LR}(k) c_{RL} | L R \rangle
+ T_{LR}( k^\prime ) c_{LR} | R L \rangle ] | k_L k_L^\prime \rangle
\nonumber \\
&& + | R R \rangle
| {\tilde \phi}_{LR}^{( k k^\prime )} \rangle ,
\label{eqn:Phi-kkprime}
\end{eqnarray}
where $ | {\tilde \phi}_{LR}^{( k k^\prime )} \rangle
= T_{RR}( k^\prime )  c_{LR} | k_L k_R^\prime \rangle
+ T_{RR} (k) c_{RL} | k_R k_L^\prime \rangle $.
Then, the fidelity for the unconditional operation
of entanglement transfer is evaluated
by tracing over the photon states and the environment
denoted by $ | 0 \rangle \langle 0 | $.
For any choice of the initial state
it is calculated to be bounded as
$ F ({\rm p} \rightarrow {\rm a}) \geq
| [ T_{LR} ]_f |^2 = F_{\rm swap} F_{\rm qm} $,
which approaches unity for the specific case
of $ g_L (k) = \pm g_R (k) $ with $ e^{i \phi_s ( k_c )} = - 1 $
in the strong coupling limit.
Furthermore, we can make actively the photon detection
$ \Pi ( k_L ) \otimes \Pi ( k_L^\prime ) $
on the ouptput state in Eq. (\ref{eqn:Phi-kkprime}),
so that the trade-off of the success probability
is made to obtain the high fidelity.
Then, the transfer of the photonic entanglement
to the atomic memories is implemented
almost faithfully even for $ \lambda_{L,R} \sim \kappa \sim \gamma $,
and the fidelity is calculated to be the same
as the 1-qubit memory,
\begin{eqnarray}
F_{\mbox{\scriptsize{ent-tr}}} = F_{\rm qm} \simeq 1 \
( \kappa_{\rm p} \ll \kappa ) .
\end{eqnarray}
The entanglement stored in the pair of atomic quantum memories is
alternatively retrieved by injecting single-photon pulses
(may be separable) to the atomic memories.
In a feasible experiment for this entanglement transfer,
a type II down-conversion light can be used
as the input polarization-entangled photonic qubit.

In summary, we have investigated a scheme of atomic quantum memory
to store photonic qubits in cavity QED.
Specifically for a $ \Lambda $-type atom with equal but opposite
dipole matrix elements, which is trapped inside an optical cavity,
the quantum-state swapping between a single-photon pulse and the atom
is ideally realized via scattering in the strong coupling cavity regime
We have derived a simple formula for calculating the fidelity
of this atom-photon swapping for quantum memory.
We have further proposed a feasible method
to implements conditionally the quantum memory operation
with the fidelity of almost unity
even for not so strong atom-photon couplings,
which is applicable for a general $ \Lambda $-type atom
with degenerate ground states.
This method can also be applied to store an photonic entanglement
in spatially separated atomic quantum memories.

The authors would like to thank M. Kitano, A. Kitagawa, and K. Ogure
for valuable suggestions and comments.
This work has been supported
by International Communications Foundation (ICF).

\end{document}